\documentstyle[aps,prl,twocolumn]{revtex}

\begin{document}
\title{\hfill October 2,1997 \\  $\: $ \\
LOW-TEMPERATURE QUANTUM RELAXATION IN A SYSTEM OF
MAGNETIC NANOMOLECULES}

\author{N. V. Prokof'ev$^{1,3}$ and P. C. E. Stamp$^{2,4}$}
\address{$^{1}$ Russian Research Center ``Kurchatov Institute",
123182 Moscow,Russia}

\address{$^{2}$ Physics Department and Canadian Institute for
Advanced Research, University of British Columbia, 6224 Agricultural
Road, Vancouver B.C., V6T 1Z1, Canada}

\address{$^{3}$ Laboratoire de Magnetisme Louis Neel, CNRS, BP 166,
38042 Grenoble, Cedex 9, France}

\address{$^{4}$ Max Planck Institut-LCMI, 25 Avenue des Martyrs,
38042 Grenoble, France}
\maketitle

\begin{abstract} We argue that to explain recent resonant
tunneling experiments on crystals of  Mn$_{12}$ and Fe$_8$,
particularly in the low-T limit, one must invoke {\it dynamic}
nuclear spin and dipolar interactions. We show the low-$T$, short-time
relaxation will then have a $\sqrt{t/\tau }$ form, where  $\tau $
depends on the nuclear $T_2$, on the tunneling matrix element 
$\Delta_{10}$ between the two lowest levels, and on 
the initial distribution of internal fields in the
sample, which depends very strongly on sample shape. The results are 
directly applicable to the $Fe_8$ system.
We also give some results for the long-time
relaxation.
\end{abstract}

\bigskip

{\bf 1.} Recent experiments on magnetic relaxation of molecular
crystals of Mn$_{12}$ and Fe$_8$ have found strong evidence for
quantum tunneling-mediated relaxation at low temperatures
\cite{Paulsen,Novak,Barbara,Friedman,Thomas,Hernandez,Sangregorio}.
This evidence comes from striking resonances observed in the
relaxation rate at certain values of applied external magnetic field
$H$ (when the energy levels of magnetic states on opposite sides of
the potential barrier match each other). These resonances exist both
in the low-$T$ limit (when only the two lowest electronic energy levels
of each molecule are involved), and also in the thermally activated
regime, when tunneling clearly takes place between higher levels. In
the Fe$_8$ system, the low-$T$ relaxation rate peak at $H = 0$ is
over 4 orders of magnitude above the rate at $H = 0.1T$, off
resonance!

In these systems the anisotropic potential acting on the molecular
"giant spins" (of spin quantum number $S$) is dominated by a strong
parabolic easy-axis term, of form $^{\parallel}{\cal H}_2^{(0)}=
-({\cal D}/S) S_z^2$; tunneling is caused by weak transverse
perturbations on this.  For Fe$_8$, ${\cal D} \sim 0.27~K$, whereas in
Mn$_{12}$, ${\cal D} \sim 0.61~K$; and $S=10$ for both molecules
\cite{Paulsen,Novak,Barbara,Friedman,Thomas,Hernandez,Sangregorio}.
Here we concentrate on the relaxation at low T (below $T_c \sim
2.2~K$ and $\sim 0.4~K$, for Mn$_{12}$ and Fe$_8$ respectively
\cite{comment1}), near $H=0$, so that only the two lowest levels 
$|10 \rangle$ and $|-10 \rangle$ of 
$^{\parallel}{\cal H}_2^{(0)}$ are involved - they are coupled by a
tunneling matrix element $\Delta_{10}$, which can in principle be
calculated \cite{Hemmen}, but only if all relevant transverse
couplings are known \cite{comment2}. Present estimates range over
several orders of magnitude, but it seems unlikely that
$\Delta_{10}$ exceeds $10^{-8}~K$ for either system - here
we will treat it as an independent parameter. 

The low-T, low-H limit is particularly interesting because of the
following simple reasoning. Recall first that a standard phase space
argument \cite{Abragam} tells us that the phonon-mediated spin
relaxation must go like $\tau^{-1} \sim \xi^3[2N(\xi) + 1]$, where
$\xi = |{\cal E}_S -{\cal E}_{-S}|$ is the bias between the two lowest
levels in an applied field, and $N(\xi)$ is the Bose function. 
If we ignore the hyperfine and dipolar
fields, $\xi = 2g\mu_BS|H|$, and we get $\tau^{-1} \sim |H|^3$;
however it is crucial to understand that even if we include these
fields, this will not change the prediction of a minimum in
$\tau^{-1}$ around $H = 0$, {\it unless} we include their {\it
dynamics}.  This is because the typical bias caused by dipolar fields
alone is $\xi_{Dip} \sim 0.1-0.5~K$; with $\Delta_{10}$ in the
nanokelvin range the effect of a distribution of {\it static} hyperfine and
dipolar fields in the sample will be simply to force almost all
molecules off resonance (resonant tunneling requiring that $\xi <
\Delta_{10}$, in the low-$T$ limit). The only way then for the molecules
to relax is then via spin-phonon interactions, and the effect of the
spread in static fields is simply to smear the minimum around $H=0$.
Notice that this argument holds even if the static fields have a
transverse component, unless this is $\sim 5T$ (enough to raise
$\Delta_{10}$ to the experimental resonance linewidth in energy, ie.,
$\sim 0.1-1K$ for the Mn$_{12}$ and Fe$_8$ systems \cite{5T}); this enormous
value is $\sim 10^2$ times the actual hyperfine/dipolar field
combination!  Thus the low-field, low-$T$ experiments force us to
consider the dynamics of the hyperfine and dipolar fields, which, by
varying the bias at each molecular site {\it in time}, is capable of
continually bringing more molecules to resonance.

At first glance, in the low-$T$ limit only the dynamic nuclear 
fields (ie., hyperfine and nuclear dipolar fields) 
can then play a role in the relaxation- once $kT \ll {\cal D}$, 
all dipolar flip-flop processes are frozen out \cite{flfl}. However we shall
see that although one needs the rapidly-fluctuating hyperfine
field to bring molecules initially into resonance, after this the
gradual adjustment of the dipolar fields across the sample, caused by
tunneling relaxation, is sufficient to bring a steady further supply
of molecules into resonance, and allow continuous relaxation. This
process is particularly important in Fe$_8$, where the hyperfine
couplings are very small. The fluctuating nuclear spin field also makes
the tunneling {\it incoherent}.  One can then write down a {\it
classical} kinetic equation for the magnetisation, whose solution
at short time is found below to have a square root behaviour, 
for almost any sample shape (although the characteristic time in the decay
depends strongly on the shape).

We shall thus find that not only do the low-T, low-H experimental results
force the inclusion of the {\it dynamics} of the internal fields into
the theory - this dynamics also leads to a characteristic (and experimentally 
testable) prediction for
the form of the relaxation.

\bigskip

{\bf 2.} We will treat the problem using a Hamiltonian 
\begin{eqnarray}
H= &\: &  {1 \over 2} \sum_{ij} V_{ij}^{(d)} \tau_z^{(i)}\tau_z^{(j)}
+ \sum_i \Delta_{10} \tau_x^{(i)} \nonumber \\
&+ & \sum_{ik} V^{(N)}(\tau_z^{(i)}, \vec{I}_k) + H^{NN}\;,
\label{Hamiltonian}
\end{eqnarray}
where the first term describes the dipolar-dipolar interactions
between molecules, the second describes tunneling, the third couples
magnetic molecules to nuclear spins $\{ \vec{I}_k \}$, and the last
term describes interactions between the nuclear spins.  This is an
effective Hamiltonian operating in the subspace of the two lowest
levels of each molecule; we choose the basis set to be $\vert S_z=
\pm S \rangle $; $\tau_z$ and $\tau_x$ are Pauli matrices, and $\{i\}$,
$\{ j \}$ label molecular sites.

We have ignored the transverse part of the dipolar coupling, since it
can only renormalise $\Delta_{10}$ in an unmeasureable way - all
flip-flop transitions to states with $S_z \ne \pm S$ are frozen out
at low $T$. Nuclear spin effects are in principle more
subtle - however, since the dipolar fields in ({\ref{Hamiltonian}) are
diagonal and not dynamic unless some of the molecules flip, one has
only to understand the dynamics of individual molecules, coupled to the
nuclear bath, assuming that
dipolar fields are frozen. This problem was solved in Ref.
\cite{our-JLTP}, sec. 4. If the nuclear $T_1$
is long (which it will be at these temperatures, since it is driven
by dipolar flip-flop processes), then the hyperfine bias field acting
on a given molecule rapidly fluctuates at a rate $T_2^{-1}$, and over
an energy scale $\Gamma_2$ which is also roughly $\sim T_2^{-1}$.
Typically $T_2^{-1} \sim 10^{-7}-10^{-5}K$, so that we expect
$\Delta_{10} \ll T_2^{-1}$. Thus at short times we can write the bias
$\xi_{j}(t)$, at molecular site $j$, as $\xi_j(t)=\xi_j+\delta
\xi_j(t)$, where $\xi_j$ results from the sum of the quasi-static
dipolar and hyperfine fields, with only a small rapidly fluctuating
component $\delta \xi_j(t)$, which nevertheless sweeps over a bias
range much larger than $\Delta_{10}$. One then finds \cite{our-JLTP}
that a molecule in quasi-static bias $\xi$ relaxes {\it incoherently}
at a rate
\begin{equation}
\tau^{-1}_N(\xi) \approx
\tau^{-1}_0 e^{-\vert \xi \vert /\xi_o} \; .
\label{tauN2}
\end{equation}
\begin{equation}
\tau^{-1}_N(\xi =0)\equiv \tau^{-1}_0 \approx {2 \Delta_{10}^2 \over
\pi^{1/2} \Gamma_2 }\; .
\label{tauN}
\end{equation}
The parameter $\xi_0$ depends on the average number $\lambda$ of nuclear
spins which co-flip with $S$. If $\lambda < 1$,
then $\xi_0 \sim \Gamma_2$; in the opposite limit
$\xi_0 \sim \lambda \vert V^{(N)}_{ki} \vert $. 
For both Fe$_8$ and Mn$_{12}$, $\lambda <1$ is most likely \cite{lambda};
in any case,
$\xi_0 \ll E_D$, where $E_D$ is the total dipolar coupling from
nearest neighbour molecules, and the exact value of $\xi_0$ will not
be too important. 

We now define a normalised 1-molecule distribution function 
$P_{\alpha } (\xi ,\vec{r}; t)$, with  
$\sum_\alpha \int d\xi \int d\vec{r} P_{\alpha } (\xi ,\vec{r} ; t ) =1$. 
It gives the
probability of finding a molecule at position $\vec{r}$, with
polarisation $\alpha =\pm 1$ (ie., in state $\vert S_z= \pm S
\rangle$), having a bias energy $\xi$, at time $t$. Molecules having
bias energy $\xi$ undergo transitions between  $\vert S_z= S\rangle$
and $\vert S_z= -S \rangle$ at a rate given by (\ref{tauN2}). 
Flipping these molecules then changes the
dipolar fields acting on the whole ensemble, bringing 
more molecules into near (or away from) resonance, 
and leading to a self-consistent
evolution of $P_{\alpha } (\xi )$ in time. The general solution of
this problem requires a kinetic equation for $P_{\alpha}(\xi
,\vec{r}; t)$.

\bigskip

{\bf 3.}  To derive a kinetic equation for $P_{\alpha } (\xi ,\vec{r}
; t ) $, we again assume that dipolar and hyperfine fields are frozen
(apart from the nuclear $T_2$ fluctuations just discussed), {\it unless} a
molecule flips. All kinetics then come from these flips, along with
the resulting adjustment of the dipolar field. We may then derive a
kinetic equation in the usual way \cite{16}, by considering the
change in $P_\alpha $ in a time $\delta t$, caused by molecular
flips, at the rate $\tau_N^{-1}(\xi )$, around the sample. This yields
\begin{eqnarray}
 \dot{P}_{\alpha } (\xi ,\vec{r} )= & &
 - \tau_N^{-1}(\xi ) [P_{\alpha } (\xi ,\vec{r} )
-P_{-\alpha } (\xi ,\vec{r} ) ] \nonumber \\
& -& \sum_{\alpha '} \int {d\vec{r}\: ' \over \Omega_0 } \int 
{ d\xi ' \over  \tau_N (\xi ' )} 
\bigg[ P_{\alpha \alpha '}^{(2)} (\xi  , \xi ';\vec{r},\vec{r}\: ')
\nonumber \\
& - & P_{\alpha \alpha '}^{(2)} 
(\xi -\alpha \alpha ' V_D(\vec{r} -\vec{r}\: ')
, \xi ';\vec{r},\vec{r}\: ') \bigg] \;,
\label{kinetic}
\end{eqnarray}
where $P_{\alpha \alpha '}^{(2)} (\xi , \xi ';\vec{r},\vec{r}\: '; t)$
is the two-molecule distribution, giving the normalized joint
probability of finding a molecule at site $\vec{r}$, in state $\vert
\alpha \rangle $ and in a bias $\xi$, whilst another is at
$\vec{r}\: '$,  in state $\vert \alpha ' \rangle $, and in a bias $\xi
'$. $P^{(2)}$ is linked to higher multi-molecule distributions by a
BBGKY-like hierarchy of equations \cite{16} . The 
first term on the right-hand
side of (\ref{kinetic}) describes the local tunneling relaxation, and
the second non-local term (analogous to a collision integral) comes
from the change in the dipolar field at $\vec{r}$, caused by a
molecular flip at $\vec{r}\: '$; the dipolar interaction $V_D(\vec{r} )
= E_D [1-3\cos^2\theta ] \Omega_0/r^3$, where $\Omega_0$ is the
volume of the unit molecular cell, and $\int d\vec{r}\: '$ integrates
over the sample volume.

We will assume that at $t=0$ the sample is fully polarized; the
initial relaxation can then be treated in a dilute solution
approximation for the fraction $(1-M)/2 \ll 1$ of flipped molecules
(where $M= \int d\xi \int (d\vec{r}/ \Omega_0 ) [P_+(\xi , \vec{r} )-
P_-(\xi , \vec{r} )] \equiv \int d\xi \int (d\vec{r}/ \Omega_0 )
M(\xi , \vec{r} )$). The bimolecular distribution function $P^{(2)}$
factorizes in this limit, ie., $P^{(2)}(1,2) =P(1)P(2)$. The simplest case
one may study is that of an {\it ellipsoidal} sample, where the
macroscopic demagnetization field is uniform. The field
distribution around randomly placed dipoles is well-known \cite{Anderson}
to be a Lorentzian up to a high-energy cutoff defined by $E_D$
\begin{eqnarray}
P_\alpha (\xi ) & =& {1+\alpha M(t) \over 2}
~ {\Gamma_d(t) / \pi \over [\xi - \alpha E(t) ]^2
+\Gamma_d^2 (t) } \;; \nonumber \\
\Gamma_d(t) & =& {4\pi^2 \over 3^{5/2} } E_D (1-M(t))  \; ;
\label{Lorentz1} \\
E(t) & = & cE_D (1-M(t)) \;,
\label{Lorentz2}
\end{eqnarray}
where $c$ is a sample shape dependent coefficient \cite{demag}, and $E(t)$ 
is the time-dependent internal field. Then
(\ref{kinetic}) gives 
\begin{equation}
\dot{M}(t) = -M(t){ 2 \over \tau_0  }
\int d\xi e^{ - \vert \xi \vert /\xi_0 }
{\Gamma_d(t) / \pi \over [\xi - E(t) ]^2 +\Gamma_d^2 (t) } \;.
\label{dilute}
\end{equation}
At very short times $t < \tau_0 \xi_0 /E_D$ 
this gives a linear relaxation 
$M(t) = 1- 2t/\tau_0$, unobservable because $\xi_0 / E_D \ll 1$. 
For $t \gg \tau_0 \xi_0 /E_D$ one gets 
\begin{eqnarray}
\dot{M}(t)& = &
-{1 \over 2\tau_{short} } \: { M(t) \over 1-M(t) } \;;
\label{tau-short}  \\
\tau_{short}^{-1}  &=& {\xi_0 \over E_D \tau_0 }\:
{32\pi  \over 3^{5/2} (c^2+16\pi^2/3^5 ) } \;.
\label{tau-sh}
\end{eqnarray}
Since (\ref{dilute}) itself is only valid when  
$1-M(t) \ll 1$,  we simply write 
\begin{equation}
M(t) \approx 1-\sqrt{t/\tau_{short} }
\;; \;\;\;\; \left( {E_D \over \xi_0}  > {t \over  \tau_0} >
{ \xi_0 \over E_D }  \right) \;.
\label{dilute-long}
\end{equation}

The square-root behaviour will be observable experimentally over a wide
time range, since $E_D/\xi_0 \gg 1$. Note also that
$\tau_{short}$ is sample shape dependent even assuming a {\it
homogeneous} demagnetisation field. If the sample is not
ellipsoidal, than the above analysis is not correct because the
problem becomes essentially inhomogeneous. We then return to
the kinetic equation (\ref{kinetic}), and notice that
if the demagnetisation varies on a length scale much greater than
the average distance between flipped spins, then (\ref{dilute})
is simply modified to 
\begin{equation}
\dot{M}(\vec{r} ,t) = -M(\vec{r} ,t){ 2 \over \tau_0  }
\int {d\xi \over \pi } 
{\Gamma_d(\vec{r} ,t)  e^{ - \vert \xi \vert /\xi_0 }
\over [\xi - E(\vec{r} ,t) ]^2 + \Gamma_d^2 (\vec{r} ,t) } \;.
\label{dilute-r}
\end{equation}
where $\Gamma_d(\vec{r} ,t)$ and $E(\vec{r} ,t)$ are defined in terms
of $M(\vec{r} ,t)$ analogously to (\ref{Lorentz1})
and(\ref{Lorentz2}); the solution is then identical to
(\ref{dilute-long}) except that $\tau_{short}$ is modified to
\begin{equation}
(\tau_{short}^{(inh)})^{-1} \sim  \xi_0 N(0) 
\tau_{short}^{-1} \;, 
\label{tau-short-r}
\end{equation}
where $N(0) = \int d \vec{r} \sum_{\alpha} P_{\alpha}
(\xi=0 , \vec{r}; t = 0)$
is the {\it initial} "density of states" for the dipolar field distribution, 
integrated over the whole sample, at bias $\xi = 0$; typically $N(0) \sim
 1 / E_{Dm}$, where $E_{Dm}$ is the average demagnetization field.

In order to verify these results, and to see when the 
square-root behaviour breaks down, we have performed Monte Carlo (MC)
simulations of the relaxation for various sample geometries, by the
usual procedure - during each time interval $\delta t \ll \tau_0 $ one
flips molecules with probability $1-\exp \{ -\delta t
/\tau_N (\xi ) \} $ and then recalculates the dipolar field distribution,
now altered by the flipped molecules (cf. Fig.1). 
The system size we can simulate is
not really macroscopic \cite{macro}, but finite 
size corrections clearly do not change 
the two main predictions coming from eqtns (\ref{tau-sh})-(\ref{tau-short-r}),
viz., (i) universality of the
square-root  relaxation at short times and (ii) the characteristic 
dependence in (\ref{tau-short-r}) of the
relaxation time on sample geometry. Clearly, the fastest
relaxation will be observed in nearly-ellipsoidal samples.

It will thus be very interesting to check in the low-T limit for this
square-root relaxation,
using {\it different sample shapes}. Our calculations are most immediately 
applicable to the Fe$_8$ system \cite{Sangregorio}, where the 
field distribution is almost
entirely due to dipolar spread \cite{Mnalso}. Confirmation of our predictions  
would then provide strong
evidence for the dynamic relaxation mechanism discussed here.
We emphasize that at higher T we do
{\it not} expect $\sqrt{t}$ relaxation, since then dipolar
flip-flop processes interfere, and $T_1$ becomes short
\cite{comment3}; moreover, the magnetisation reversal also proceeds via higher
levels, through mixed activation/tunneling processes.
Coupling to the
phonon bath is then crucial, which essentially changes the theory. 

Finally, we consider the relaxation
when $t \gg \tau_{short}$. This problem is greatly complicated by the
development of intermolecular correlations in $P^{(2)}$, $P^{(3)}$,
etc., so that one may no longer factorize them. However one way of
avoiding this {\it experimentally} would be to let the system
substantially relax at high T, then cool to low
T; one would then be in the long-time relaxation regime,
but with initial condition arranged to give a factorizable
$P^{(2)}$. In this limit another analytic solution for the homogeneous
(ie., ellipsoidal) case can be found from (\ref{kinetic}), when $M \ll 1$ and
$P^{(2)}_{\alpha \alpha '}(\xi , \xi '; \vec{r},\vec{r}\: ') = 
P_{\alpha }(\xi )P_{\alpha '}(\xi ')$; one finds {\it exponential}
relaxation, at a rate 
\begin{equation}
\tau_{long}{-1}\approx { 2\xi_0 \over E_{max} \tau_0 [1+\kappa \ln
(E_{max} /\pi \xi_0 ) ] } \;, 
\label{tau-long}
\end{equation}
where $\kappa \sim 1$ is a numerical coefficient, and $E_{max}$ is the
spread in dipolar fields in this nearly depolarized limit. Details of
the derivation 
will be given in a longer paper \cite{19}. 

We would like to thank B.Barbara, T.Ohm, L.Thomas, and C.Paulsen for
discussion of their experiments. 
This work was supported by the Russian Foundation for Basic Research
(grant 97-02-16548), and by NSERC and the CIAR in Canada.

\bigskip
{\bf Figure 1}: (a) Monte Carlo (MC) simulations of the relaxation in 
two samples, each made from a cubic lattice of molecules. In (a) the 
relaxation is shown as a function of $\sqrt{t/\tau_0 }$ 
for (1) a cubic sample of 
$(50)^3$ molecules, and (2) for a sphere of diameter 50 lattice spacings 
((b)  shows the same relaxation as a function of $t/\tau_0 $). 
The dashed line shows $\sqrt{t}$ behaviour; 
the sample relaxation shows multimolecule correlation effects 
once $M(t) \lesssim 0.93$.

{\bf Figure 2} The density of states $N(\xi)$ for the distribution 
of bias fields,
integrated over the sample (cf. text) is shown for time $t=0$ (where 
finite size effects smear the zero energy delta function), and for $t = 
0.1 \tau_0$, for the spherical sample. Energy (and density of states) 
use units where $\xi_0 = 1$ and $E_D =20$. 
The fraction of states in the 
resonant window of width $\xi_0$ around zero energy, at $t=0$, was $0.79$
(sphere), and $0.037$ (cube); the ratio 
$\sqrt{0.79 / 0.036 } \approx 4.6$ 
corresponds fairly well to the ratio $\sim 4$ between 
the straight-line slopes in (a).


\begin{thebibliography}{99}

\bibitem{Paulsen}  C. Paulsen and J.G. Park, in {\it "Quantum Tunneling of
Magnetisation-QTM'94"} (ed. L. Gunther and B. Barbara), Kluwer 
Publishing, pp. 189-207
(1995).

\bibitem{Novak}  M. Novak and R. Sessoli, pp. 171-188, in ref. 1. 

\bibitem{Barbara}  B. Barbara {\it et al.}, JMMM, {\bf 140-144}, 1825 (1995).

\bibitem{Friedman}  J.R. Friedman {\it et al.}, Phys. Rev. Lett., {\bf 76},
3830-3833 (1996).

\bibitem{Thomas}  L. Thomas {\it et al.}, Nature {\bf 383}, 145-147 (1996).

\bibitem{Hernandez}  J.M. Hernandez {\it et al.}, Europhys. Lett., {\bf 35},
301-306 (1996).

\bibitem{Sangregorio}  C. Sangregorio, T. Ohm, C. Paulsen,
R. Sessoli, and D. Gatteschi, Phys. Rev. Lett., {\bf 78}, 4645 (1997).

\bibitem{comment1} These numbers are taken from the experiments 
\cite{Thomas,Sangregorio}; the crossover is clear from the
T-independent relaxation rate below $T_c$.

\bibitem{Hemmen} See, e.g., L. van Hemmen and A.Suto, Physica B {\bf
141}, 37 (1986).

\bibitem{comment2} Note that even very small higher-order transverse
couplings (up to the 20th-order in $S^+$ and $S^-$) can make
important contributions to $\Delta_{10}$, simply because lower-order
couplings contribute to $\Delta_{10}$ with large exponents (e.g., the
transverse coupling $^{\perp}{\cal H}_0^{(2)} =E(S_x^2-S_y^2)$ gives
a contribution $\sim E (E/2{\cal D} )^9$), thereby strongly
suppressing their effect on $\Delta_{10}$. This makes $\Delta_{10}$
impossible to calculate, since such higher-order couplings are
unmeasurable.

\bibitem{Abragam}  A. Abragam and A. Bleaney, {\sl Electron Paramagnetic
Resonance of Transition Ions}, Clarendon (1970).

\bibitem{5T} Recall that the effect on $\Delta_{10}$ from a static transverse
$H$ is $\sim g\mu_BSH_{\perp}(g\mu_BSH_{\perp}/{\cal D})^{19}$, and that for
the Mn$_{12}$ and Fe$_8$ systems, $({\cal D}/g\mu_BS) \sim 10T$.

\bibitem{flfl} The dipolar 
flip-flop transitions at low $T$ go at a rate 
$\lambda_{fl} \sim 
\Omega_{dip} \exp \{ -({\cal E}_9 - {\cal E}_{10})/k_B T \}$, where 
$\Omega_{dip} \sim 10^6 - 10^8 Hz$. Note that the {\it direct} 
effect of flip-flop 
processes on the kinetic equation is very small (the concentration
of molecules in state $|9 \rangle$ is $\sim 
\exp \{ -({\cal E}_9 - {\cal E}_{10})/k_B T \}$); they are 
dangerous only because they drive nuclear $T_1$ processes, which can sweep
$\xi$ over a much larger range than $\xi_0$.

\bibitem{our-JLTP}  N.V. Prokof'ev, P.C.E. Stamp, J. Low Temp. Phys.{\bf 104%
}, 143 (1996); and see also pp. 347-369 in ref. 1.

\bibitem{lambda} In the simplest case where all nuclear spin effects 
come via the hyperfine coupling $\sum_{k} \omega_k {\vec s}_k .
{\vec I}_k$, summed over the electronic spins in the molecule, one has
$\lambda \sim \sum_{k} (\omega_k / {\cal D})^2 \ll 1$ (cf. ref.
\cite{our-JLTP}), for both the Mn$_{12}$ and Fe$_8$ systems. 
Nuclear dipole-dipole interactions will change this estimate, but not 
drastically.


\bibitem{16} See, eg., N.G. van Kampen, {\sl Stochastic Processes in Physics
and Chemistry}, North-Holland (1981).

\bibitem{Anderson} P.W. Anderson, Phys. Rev. {\bf 82}, 342 (1951); A.
Abragam, {\it Principles of Nuclear Magnetism} (O.U.P.), Sec. IV.4,
p. 125 (1961). 

\bibitem{demag} The coefficent $c$ is defined by eqtn. \ref{Lorentz2}. For a 
prolate spheroid, 
$c=(2 \pi /3)  [a^4+a^2-3a\sqrt{a^2-1}\ln (a+\sqrt{a^2-1} ) -2 ]/(a^2-1)^2$,
where $a$ is the ratio of the longitudinal axis to its perpendicular; 
analytic formulas can be found for any ellipsoid \cite{19}.

\bibitem{macro} Even with 
$~10^5~$ molecules,  statistical fluctuations in $P_{\alpha}(\xi,t)$
inside the small resonance
window (width $\xi_0$) are large if $E_D/ \xi_0 \gtrsim 30$ 
(which correspondingly limits the timescale 
over which (\ref{dilute-long})
is observed). In real systems $E_D/\xi_0$ can be very large. 
 
\bibitem{Mnalso} In Fe$_8$, the hyperfine field is $\sim 3G$ (due mainly to 
protons), whereas in Mn$_{12}$, it is more like $250G$; in both systems, 
$E_D \sim 1000G$. Provided $T_1 \gg \tau_{short}$ (so only 
$T_2$ fluctuations matter),
the $sqrt{t}$ prediction still holds; however the extra large random hyperfine 
field in Mn$_{12}$ means that (\ref{tau-short-r}) is no 
longer precisely obeyed
\cite{19}.

\bibitem{comment3} When $T_1 \ll$ than the
experimental time-scale, {\it and } hyperfine fields are larger than
intermolecular dipolar fields, one finds exponential relaxation
at higher $T$; see N.V. Prokof'ev and P.C.E. Stamp, preprint (April
1997). 

\bibitem{19} N.V. Prokof'ev, P.C.E.Stamp, longer paper in preparation,
will discuss the long-time behaviour (including the derivation of 
(\ref{tau-long})), intermolecular correlations, and also the 
low-$T$ behaviour in 
large applied fields.
\end{thebibliography}
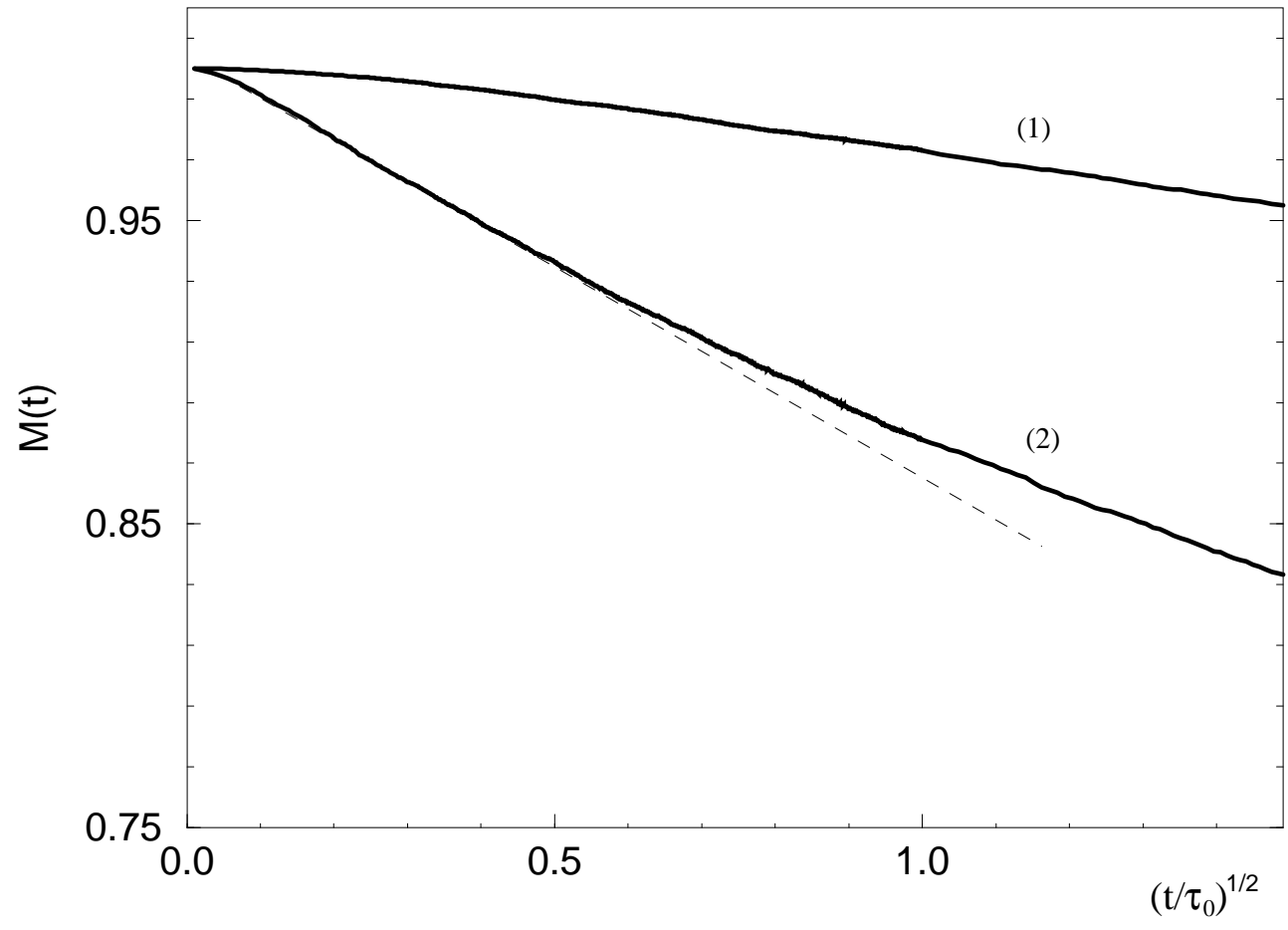
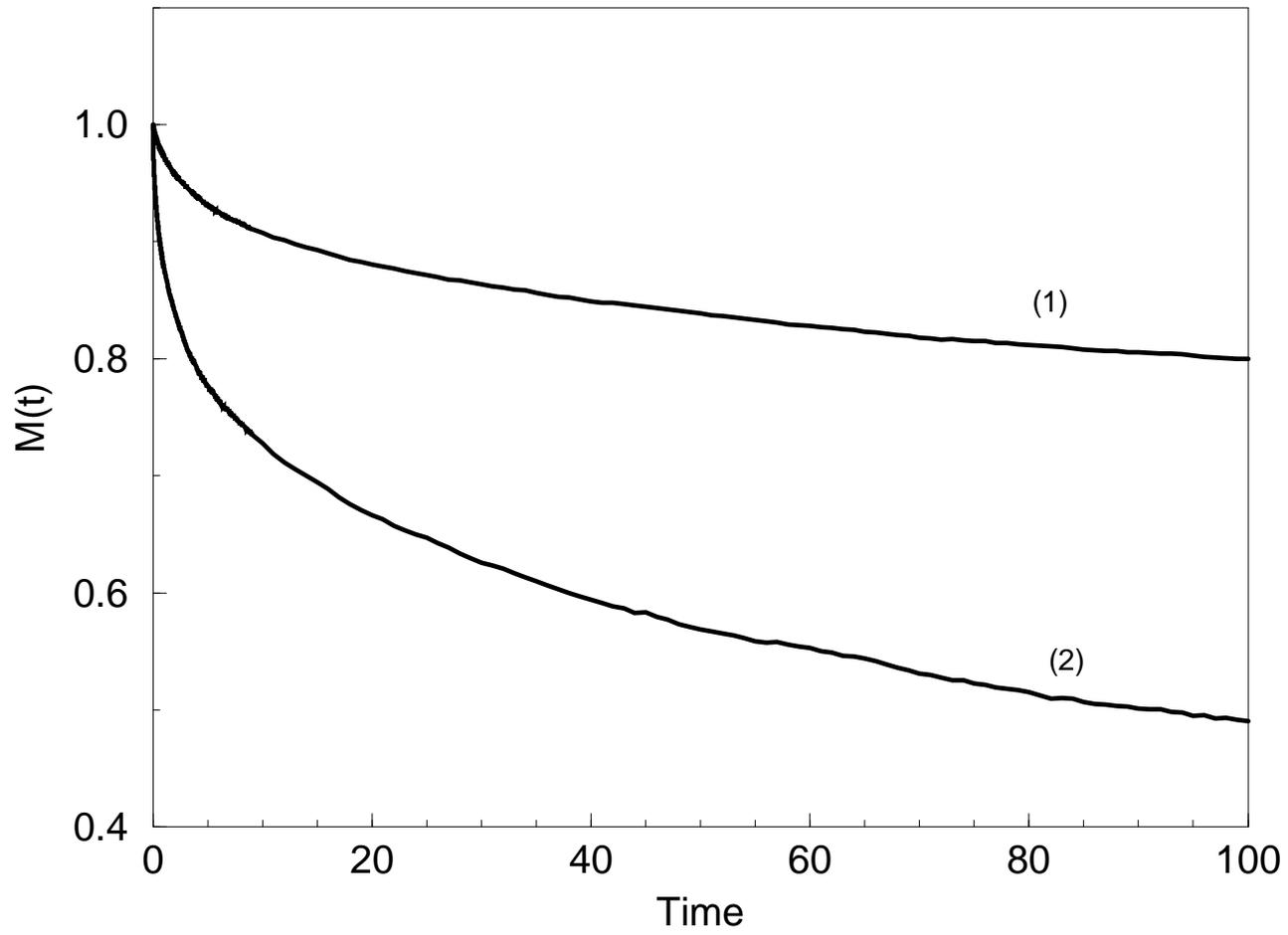
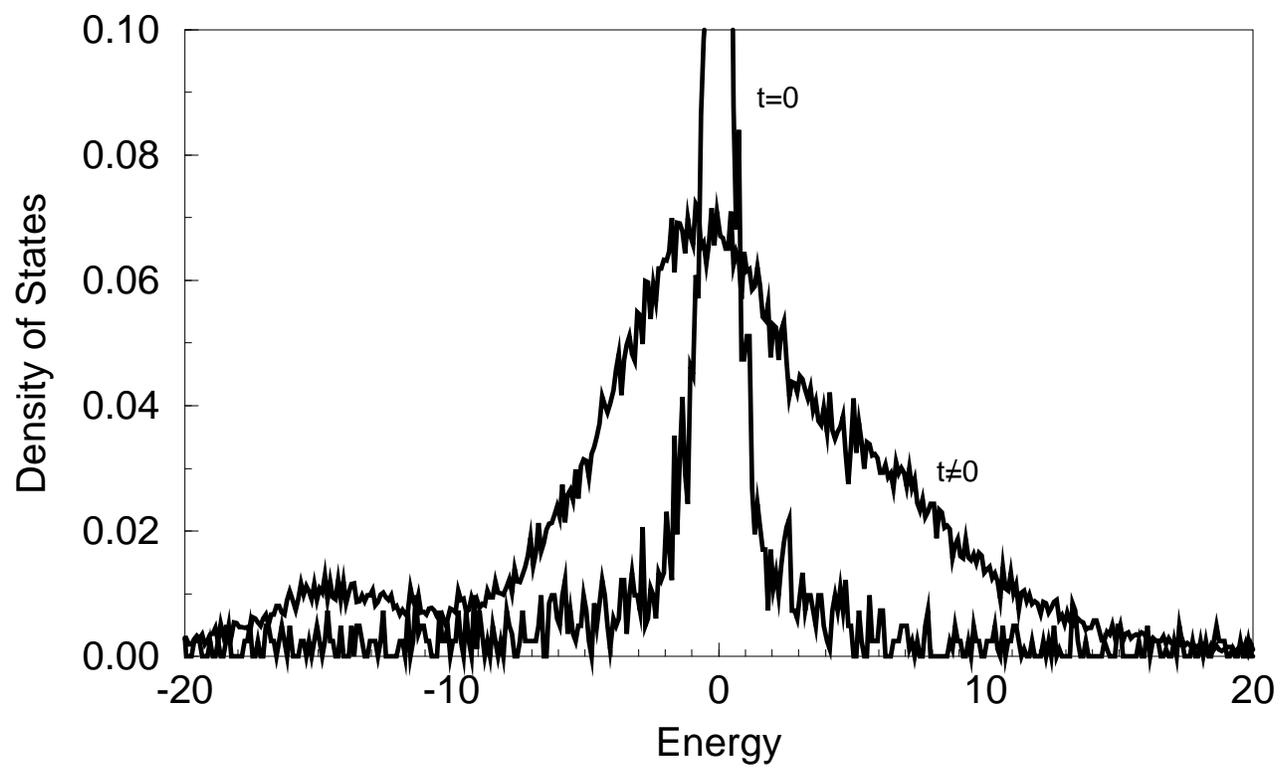
\end{document}